# Probing the Limits of the Talbot-Plateau Law

Ernest Greene and Jack Morrison
Department of Psychology
University of Southern California
Los Angeles, CA 90089

**ABSTRACT**

The Talbot-Plateau law specifies what combinations of flash frequency, duration, and intensity will yield a flicker-fused stimulus that matches the brightness of a steady stimulus. It has proven to be remarkably robust in its predictions, and here we provide addition support though the use of a contrast discrimination task. However, we also find that the visual system can register contrast when the combination of frequency and duration is relatively low. Flicker-fused letters are recognized even though they have the same physical luminance as background. We propose that this anomalous result is produced by large disparities in the duration of bright and dark components of the flash cycle, which brings about unexpected differential activation of ON and OFF retinal channels.

**SIGNIFICANCE STATEMENT**

Oscillation of retinal ON and OFF circuits appear to have a role in encoding of stimulus attributes. The ability to use frequency and the timing of flash components to modify perception should prove useful for discovering the function of these oscillations.

**INTRODUCTION**

*[W]e have only to take a white circle, with one of its sectors painted black, and make it revolve rapidly. It will appear, as everyone knows, of a uniform gray tint... In very point of the circle the white and black parts meet the eye during the same proportion of time, and therefore the tint is uniform throughout.*                                                           H.F. Talbot (1834, pg. 329)

   The goal of the research reported here is to press the limits of the Talbot-Plateau law, which involves perceived brightness of flash sequences. The Talbot-Plateau law specifies when a flash sequence will deliver light energy that makes it appear equal in brightness to a steady stimulus source. The law requires that the frequency of the flash sequence be high enough to produce fusion of individual flashes, thus providing a "flicker-fused" stimulus that appears to have steady light emission.  Additionally, the Talbot-Plateau law requires reciprocity of flash duration and flash frequency, which assures an average rate of flash energy (formally, light flux) across a range of (duration x frequency) combinations.  And finally, there is a flash intensity for a given (duration x frequency) combination that will provide flux that matches the flux of the steady stimulus, whereupon the two stimuli will be seen as being equal in brightness.

This rather improbable prediction was formulated in the mid-1800s, based on conjecture and experimental evidence provided by Talbot (1833) but also independent findings by Plateau (1830).  The earliest work used a spinning mirror that bounced a flicker-fused beam of sunlight to a spot and compared its brightness to an adjacent spot that was illuminated by a steady beam.  Light intensity was controlled by varying the size of the aperture through which the light beam passed, or through the relative alignment of polarizing filters (Talbot, 1833; Stewart, 1888).  Once the adjustments provided judgments of equal brightness for the flicker-fused and steady spots, the spin rate could be increased without changing the perceived brightness.  This established the reciprocity principle outlined above, in that faster rotation would increase the frequency of flicker and produce a compensating reduction of flash duration.

Most of the work that established the Talbot-Plateau principle as a "law" was done in the early part of the 20th century (Sherrington, 1902, 1904; Hyde, 1906; Beams, 1935; Gilmer, 1937; Bartley, 1937, 1938a, 1938b, 1939.  Many investigators replaced the spinning mirror with spinning disks that had open sectors that could more precisely determine the duration of successive flashes.  We are not aware of any tests being done using modern electronic equipment until Szilagyi (1969) tested brightness judgments of two LEDs, one that was flashing at 30 Hz and the other providing steady light emission.  He reported confirmation of the Talbot-Plateau law for a 30 Hz flicker-fused stimulus with flash durations ranging from one microsecond to ten milliseconds.  The present laboratory has recently reported additional confirming evidence using flashes from an array of LEDs (Greene & Morrison, 2023).  That work varied steady intensity levels, used flash durations ranging from one microsecond to ten milliseconds, and varied frequency as required to maintain reciprocity of duration x frequency.  The Talbot-Plateau law provided surprisingly good predictions of the flash intensity needed to achieve brightness match across more than seven orders of magnitude.

The Talbot-Plateau law is about judgments of relative brightness of two stimuli, which formally is a contrast discrimination. The two stimuli are judged as being equal in brightness when one cannot perceive any contrast. Most tests of the Talbot-Plateau predictions display two small spots of light, one providing steady light emission and the other flickering at a frequency that appears steady. The Greene & Morrison (2023) experiments demonstrated that the law is impressively accurate across a large range of flash duration and frequency combinations, but those judgments were made against a dim background. We have no assurance that this would be true if the judgments were being made against a background with higher luminance, nor have earlier studies provided sufficient insight about the matter.

To address this issue, we implemented a novel way to assess the Talbot-Plateau law by using a contrast discrimination task. Instead of placing a flicker-fused stimulus adjacent to a steady stimulus, it was displayed as a foreground area, shaped as a letter, against a steady background -- see Figure 1. For a given frequency x duration combination, letters should be visible for flash intensities that are higher or lower than what Talbot-Plateau requires for matching background brightness. But the flicker-fused letters should not be visible at or near the intensity where Talbot-Plateau predicts that the letter and background will have the same brightness level. If one scores each trial according to whether the subject is able to identify the letter, or not, we expect a U-shaped curve, as illustrated in Figure 2.

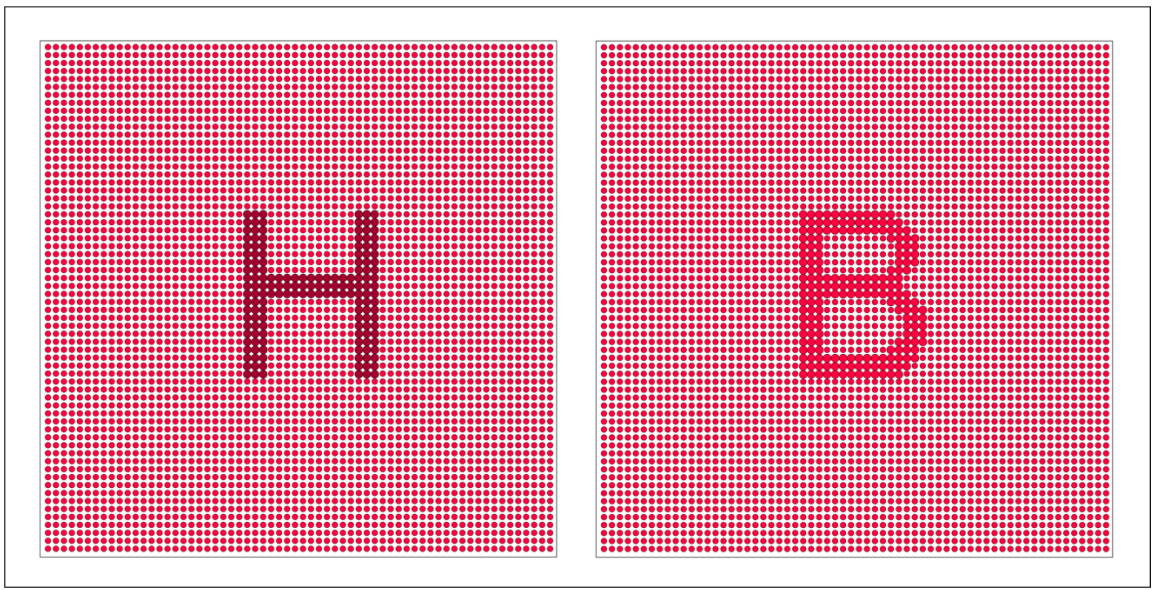

Figure 1. Flicker-fused letters can be displayed as foreground against a background of illuminated LEDs. The left panel illustrates that the letter will appear dark if the average intensity of the flicker-fused flash sequence is less than the steady background intensity. The right panel illustrates that it will appear bright if the average intensity is greater than background.

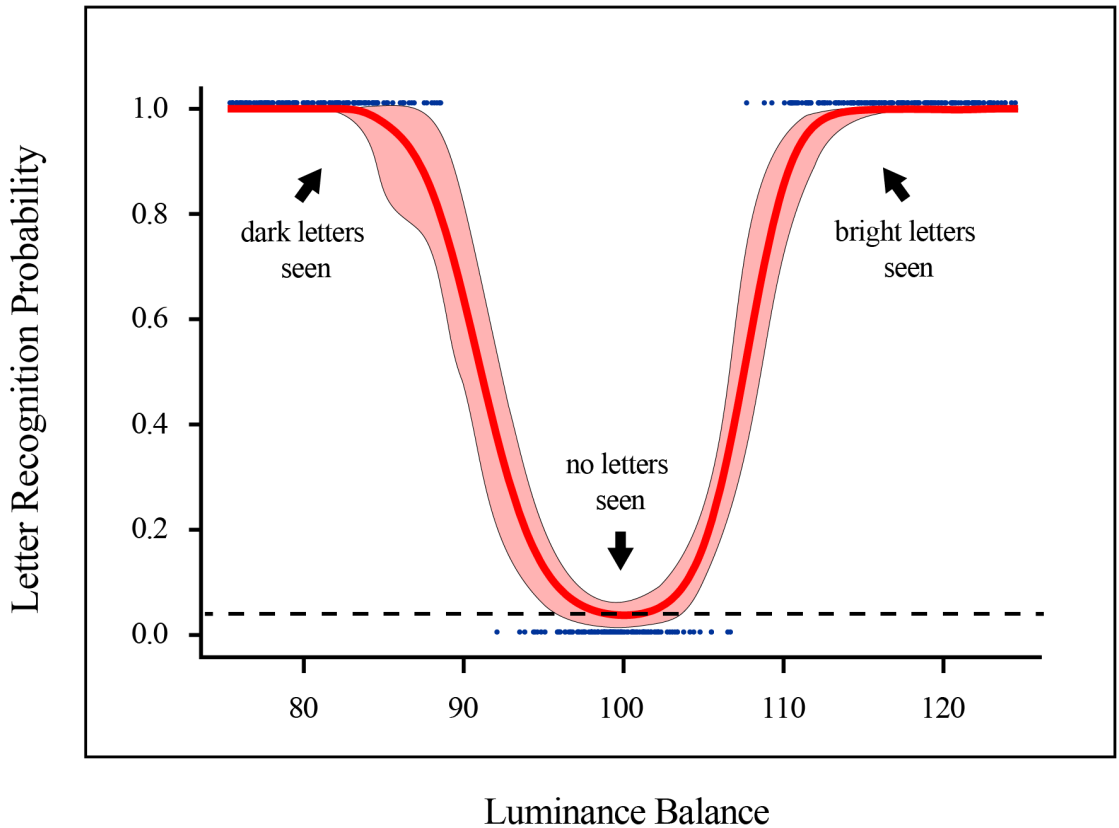

Figure 2. Present experiments displayed flicker-fused letters against a steady background and varied the degree to which their luminance levels matched. This figure illustrates the outcome that would be expected on the basis of the Talbot-Plateau law, with an abscissa value of 100 designating the condition where the letter and background have equal luminance. Letters should be recognized when the luminance levels differ substantially, but they should be invisible when the luminance levels are balanced. Dots at the top and bottom of the illustration mark intensities at which letters would be displayed, with recognition being scored as 1 and failure to see a letter being scored as 0. A logistic regression that was calculated across these data would reflect whether the Talbot-Plateau law correctly predicted the probability of letter recognition.

The findings reported here used several different treatment combinations that varied the frequency, duration, and intensity of flashes, each combination providing a range of luminance differentials. Many of the combinations produced results consistent with the prediction illustrated in Figure 2. However, there were significant effects that are at odds with the Talbot-Plateau law. For frequencies at or near 250 hz in combination with ultrabrief flash durations, respondents were able to identify flicker-fused letters that provided the same physical luminance level as the background. These findings require some form of amendment to the Talbot-Plateau law. We review possible physiological mechanisms for registering luminance and contrast and formulate an hypothesis about how channels of the retina might allow contrast to be signaled even when the stimuli provide equal luminance of foreground and background.

**METHODS**

*Display Equipment*

Experiments were conducted using an LED display board, this being a fourth-generation LED display for the present laboratory. The board was designed and fabricated by Digital Insight. Board components and additional details on the hardware are specified in Supplemental Methods. The display portion consisted of a 64 x 64 array of LEDs that have peak emission at 633 nm (red). This type of LED has a typical rise/fall time of 3 ns. The shape of the stimulus pulses (flashes) is determined by the slowest component of the driving circuit, which provided a rise/fall time estimated to be 70 ns. A thin sheet of frosted plastic covered the array to provide a small amount of light diffusion. The resulting dot had a spatial distribution that was approximately Gaussian, with 90% of the light being delivered within the span of 5.3 mm (hereafter designated to be the dot diameter). Center-to-center spacing and total span of the array (both horizontal and vertical) were 4.83 and 309 mm, respectively. The viewing distance of these experiments was 1.5 meter, so the corresponding visual angles for dot diameter, center-to-center spacing, and total span were 12, 11, and 707 minutes, respectively. Temporal resolution of all system components was in the nanosecond range. Oscilloscope traces from a fast photodiode were captured to verify the timing of the ultra-brief flashes. Minimum flash duration and the precision of flash-sequence timing was 1 microsecond (µs). A compact Windows PC controlled the system using Tcl/Tk custom applications.

*Treatment Sets*

Light intensities were specified as luminance (flux emitted from a flat surface), reported here as $Cd/m^2$. For each Treatment Set, a background intensity of 8 $Cd/m^2$ provided steady illumination that extended across the full array, except during the interval when the flicker-fused letter that was provided at the center of the array. The dots of the letter patterns were styled as Arial 33-point TrueType fonts, the same as in previous work (Greene & Visani, 2015).

The treatments described below specified frequencies and flash durations and varied flash intensities to alter the luminance of flicker-fused letters. The flash intensity at which the Talbot-Plateau law predicts luminance that is equal to the background was designated as 100%T-P. Flash intensities that were below or above that level were specified in the same manner, e.g., 75%T-P and 125%T-P, respectively. Pilot work that displayed letters at 100%T-P found that flicker could not be perceived against the 8 $Cd/m^2$ background at frequencies above 45 hz.

For Treatment Sets 1-4, flash intensity was chosen at random from within the range from 75%T-P to 125%T-P. (Details for Treatment Set 5 are provided below.) The letter and (frequency x duration) combination were chosen at random for successive trials, but requiring that each combination be displayed the same number of times. Background emission remained on throughout the test session, with each of the trials displaying the flicker-fused letter for 300 ms.

*Treatment Set 1* -- Frequency of the flicker-fused letter was tested at 250, 2,500, 25,000, and 250,000 hz. Each frequency was combined with a flash duration that would maintain a 50% duty cycle, i.e., 2000, 200, 20, and 2 µs respectively. A 50% duty cycle delivers light for half of the interval in which the flicker-fused letter is being displayed, the other half being dark, so the Talbot-Plateau principle predicts that flash intensity should be twice steady intensity for the flicker-fused stimulus to be seen as equal in brightness. Therefore, at each frequency the 100%T-P flash intensity was 16 $Cd/m^2$, thus providing an average intensity for the flicker-fused letter that matched the average background intensity of 8 $Cd/m^2$.

*Treatment Set 2* -- Frequency of the flicker-fused letter was again tested at 250, 2,500, 25,000, and 250,000 hz. A 2 µs flash duration was used in combination with each of the frequencies, which provided corresponding duty cycles of 0.05%, 0.5%, 5%, and 50%. As the duty cycle was reduced, the amplitude of the flash that would satisfy the Talbot-Plateau law increased, so a 2 µs flash at 250 hz was extremely bright.  Nonetheless, the average intensity across both bright and dark portions of the cycle still matched the steady background of 8 Cd/m$^2$.

*Treatment Set 3* --   The (250 hz x 2 μs) conditions of Treatment Set 2 provided results that were at odds with the Talbot-Plateau law.  Respondents were able to identify the flicker-fused letters even though the letters were being displayed at the same physical luminance as the background.  To better assess the role of flash frequency, Treatment Set 3 tested flicker-fused letter recognition with frequencies at 250, 500, 1000, and 2000 hz, each combined with a flash duration of 2 μs.

*Treatment Set 4* --  To assess whether anomalous naming of letters could be attributed to the brevity of the flashes, flash durations of 2, 20, 200, and 2000 μs, were each displayed at 250 hz.  One might note that this provides for corresponding duty cycles of 0.05%, 0.5%, 5%, and 50%.

*Treatment Set 5* --  To assess the role of flash brightness, flash intensities were specified as departures from a background baseline at two levels: 100% and 50%.  The 100% condition provided a flash cycle that matched the intensities of a 250 hz x 2 μs combination.  The dark portion of the flash cycle dropped from being equal to background to zero emission for 3998 μs. The 2 μs bright flash delivered enough light to yield an average intensity that matched the background across the full 4 ms flash cycle.  The 50% condition provided bright and dark departures from background that were half the amplitude of the 100% condition.  The luminance of the 50% flicker-fused letter also matched the luminance of the background, but the flash amplitude was only half the intensity of the 100% condition.

Each bright/dark flash cycle was displayed for a duration matching the period of a 250 hz frequency, i.e., 4 ms.  But instead of presenting the letters as a continuous flash sequence across an interval of time, e.g., 250 hz for 300 milliseconds, the letters were displayed for a specified number of cycles, specifically: 1, 2, 4, 8, 16, 32, 64, 128.

### *Execution of Experiments*

Experimental methods were approved by the Institutional Review Board at USC.  Twenty-five USC undergraduate respondents were paid to provide contrast discriminations.  Each respondent was provided with a description of the task and the judgments that were required.  They were informed that participation could be discontinued at any time without penalty.

Each respondent was tested and found to have 20/20 visual acuity, one requiring corrective lenses.  Respondents were tested individually.  Each sat in a room where ambient illumination was provided by a single 7 Watt DC daylight-white light bulb (6000 K). Light intensity was controlled by pulse-width modulation to maintain color temperature. The bulb was positioned on the wall 2.25 m from the floor, this being behind and above the respondent's head and therefore was not within the field of vision. Ambient light level was 10 lux, this being measured using an Extech Model LT300 light meter with the sensor facing the intersection of ceiling and back wall behind the display board. At this light level, average pupil diameter was previously found to be 6.66 mm.

The LED array was tangent to the line of sight at a distance of 1.5 m and the stimulus displays were judged with both eyes open. The successive trials of a given test session displayed flicker-fused letters against steady background intensities, as described above. On each trial of a given experiment, the computer randomly selected a letter to be displayed, and the respondent was expected to name the letter or report that no letter could be seen. The experimenter logged the response and launched the next experimental trial. Neither the respondent nor the experimenter was provided with feedback as to whether the response was correct. Each of the five respondents that were tested with a given Treatment Set judged every treatment combination provided by the set, providing judgments across 400 trials. All trials were completed by the respondent in about 50 minutes.

The use of letter recognition has benefits over a conventional contrast discrimination task. The traditional method would call for a binary choice, e.g., Yes/No, for whether the brightness level of the flicker-fused stimulus matches the steady stimulus. A threshold criterion would normally be set at 50%, which means that half of the time the person does, in fact, be seeing flicker. Although this is a perfectly legitimate operational criterion, it provides less certainty for specifying the energy at which the two zones are equally in brightness. For letter recognition, chance performance is less than 4%, this being the probability of randomly picking a given letter out of an inventory of 26. Being able to say that the brightness of the two displays has equal energy within a measurement error of 4% is better than saying that they are equal with a 50% probability. An additional benefit of letter reporting is that respondents maintain better vigilance, i.e., become less bored, and therefore provide more stable responding across the test session.

Respondents were expected to name any letters that were perceptible. They were also allowed to guess, or could answer "none" or "don't know" or by some other utterance indicate that they did not see or could not identify the letter. Correct identification of letters was scored as a 1, reported inability to see any letter was scored as a 0, and incorrect guesses were scored as 0.5.

Initial modeling of the data was done to determine whether the guesses contributed any useful information. The models did not differ significantly from those done using only the 1/0 data, so these responses were removed from the reported results to reduce ambiguity in interpretation of findings.

*Statistical Modeling*

For modeling of data from Treatment Sets 1-4, mixed-effects repeated-measures logistic regression models were used to estimate the mean recognition probabilities at various intensity, frequency (Treatment Sets 1-3), or flash duration (Treatment 4) levels. B-Spline forms of the intensity levels, frequency or flash duration levels, and their interactions were included as fixed effects. Subjects were treated as random effects. A nonparametric bootstrap method with 1000 samples was used to obtain the bootstrap percentile 95% confidence intervals (CIs) of the estimated means at each frequency level (Treatment Set 1). Results from these bootstrap samples were also used to calculate the minimum probabilities and their corresponding 95% CIs at each frequency level.

For modeling of data from Treatment Set 5, a mixed-effects repeated-measures logistic regression model were used to estimate the mean recognition probabilities and their 95% CIs at various number of flash cycles with either 50% or 100% bright/dark flash intensities. Piecewise linear forms of the flash cycles (in log2 scale), flash intensity percentages, and their interactions were included as fixed effects. Subjects were treated as random effects.

A p-value of less than 0.05 was considered significant for all statistical tests. SAS 9.4 (SAS Institute) was used for all statistical analyses.

# Results

*Treatment Set 1*   The first set of five respondents judged flicker-fused letters that were displayed at 250, 2500, 25000, and 250000 hz, with flash durations being 2000, 200, 20, and 2 microseconds, respectively. Each combination of frequency and duration provided a 50% duty cycle. The sampled flash intensities bracketed the Talbot-Plateau prediction for the level that would match the brightness of the 8 $Cd/m^2$ background. The expectation was that intensities below that level would be seen as darker than background, thus allowing recognition of the letters. Conversely, intensities above the prediction would be seen as brighter than background -- also allowing identification of the letters. Recognition probability was expected to be at chance (4% for an inventory of 26 letters) at or near a flash intensity of 100%T-P. Therefore, logistic regression models should yield a U-shaped function, with many non-recognition trials in the vicinity of the Talbot-Plateau prediction.

For Treatment Set 1, the logistic regression models for respondents manifested strong U-shaped functions, as illustrated in Fig 3. We are struck by the narrow span of the brightness-match range. At each frequency the probability of recognition transitioned from full recognition to non-recognition within an intensity range of 10-15% on each side of model minima. Also, at each frequency the minimum probability of recognition was very close to the value predicted by the Talbot-Plateau law, being shifted to be above or below 100%T-P by no more than 10%.

Minimum probability of recognition was at chance level (4%) for displays done at 2,500, 25,000, and 250,000 hz, as reflected by the overlap of the confidence intervals. Letter recognition was slightly above chance for the 250 hz condition.

*Treatment Set 2*   This task was designed to see how a change in duty cycle would affect recognition performance. A 2-microsecond flash duration was delivered in combination with frequencies at 250, 2,500, 25,000, and 250,000 hz, which provided corresponding duty cycles that ranged from 0.05% to 50%. The logistic regression models are shown in Fig 4. The models for 2500, 25,000, and 250,000 hz were very similar to those found in Treatment Set 1, but the plunge in letter recognition at 250 hz was completely absent. The recognition was expected to be at or near chance for flash intensities in the vicinity of 100%T-P. The letter should simply disappear into the background, whereupon the respondent would report that no letter was seen.

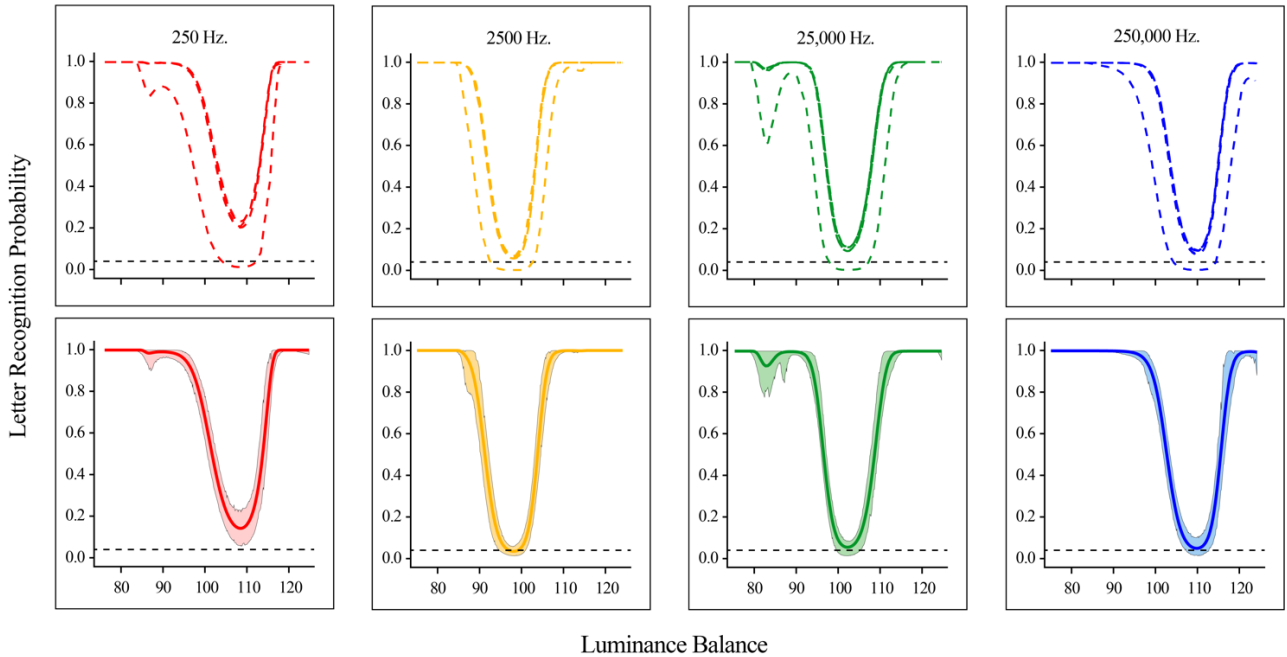

Figure 3. Respondents 1-5 judged each of the four frequencies provided by Treatment Set 1. Individual regression models are shown in the upper panels and the mean models are shown in the lower panels. Probability of letter recognition was at or near chance levels with flash intensities that produced a luminance level that matched background luminance. Fairly reliable letter recognition was found at flash intensities that were more than 10% above or below the Talbot-Plateau prediction.

But across 2000 display trials (400 trials for each of the five respondents), there were only four trials on which a "no letter" response was given. One subject reported that no letter was seen on two of the trials and two others did so on one trial each. Those four "no letter" judgments were likely produced by lapses in attention, given that the flash intensities were well away from the zone where one would expect to find brightness match.

     We should also note that minimum recognition probabilities were also above chance for the 2500 and 250,000 hz frequencies, these being at 21% and 13% respectively. Being above the chance level of 4% is at odds with the Talbot-Plateau law. In the vicinity of 100%T-P, the letters are being displayed at a luminance that equals the luminance of background. If any can be identified, it means that they are being registered by the visual system as having luminance contrast.

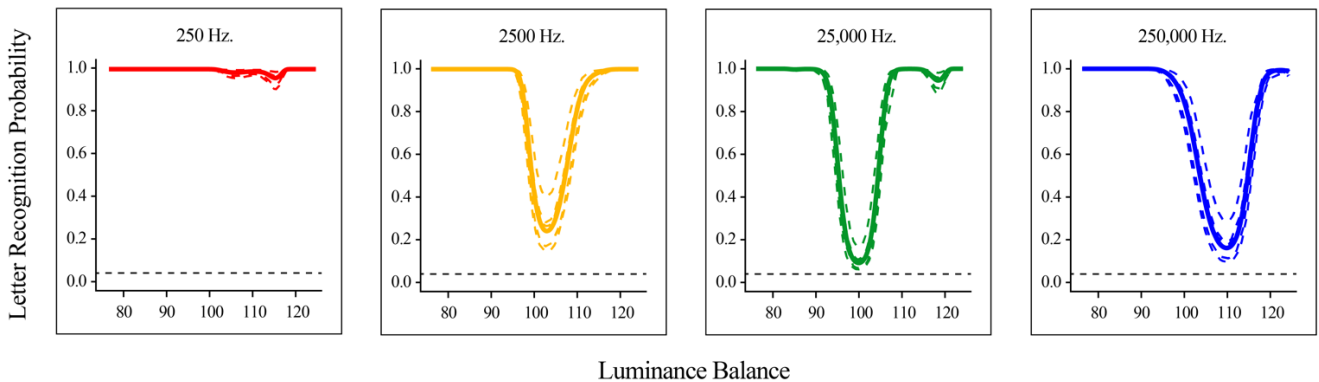

Figure 4. Regressions models for respondents 6-10 are shown for Treatment Set 2, with the mean model overlapping subject models. Non-recognition of letters was found in the vicinity of 100%T-P for 2500, 25,000, and 250,000 hz. However, letters were identified across the full range of flash intensities for the 250 hz condition. The physical luminance of a flicker-fused letter that was displayed at 250 hz with 2-microsecond flashes was equal to luminance of the background, yet luminance contrast was registered by the visual system.

*Treatment Set 3*   The probability of recognition had a minimum of 21% for displays that combined 2500 hz with 2-microsecond flash durations. Reducing the frequency to 250 hz raised the recognition probability to 100%. This suggests a possible frequency gradient for eliciting the perception of luminance contrast. Treatment Set 3 displayed the letters at 250, 500, 1000, and 2000 hz, each in combination with 2- microsecond flash durations. Logistic regression models for five new respondents (11-15) as well as the mean models are provided in Fig 5. One can see a consistent reduction of probability of recognition as the frequency is reduced from 2000 to 250 hz. A combination of ultrabrief flash duration with frequencies in the vicinity of 250 hz are providing displays wherein the flicker-flashed letter contrasts with the background. The letter is perceived even though the luminance of the letter matches the background, i.e., they are "luminance balanced."

Unlike the results in Treatment Set 2, here the 250 hz condition produced a small number of "no letter" responses and these were provided in the vicinity of 100%T-P. This indicates that the recognition of letters at 250 hz in Treatment Set 2 was not due to shifting the luminance-match zone away from 100%T-P.

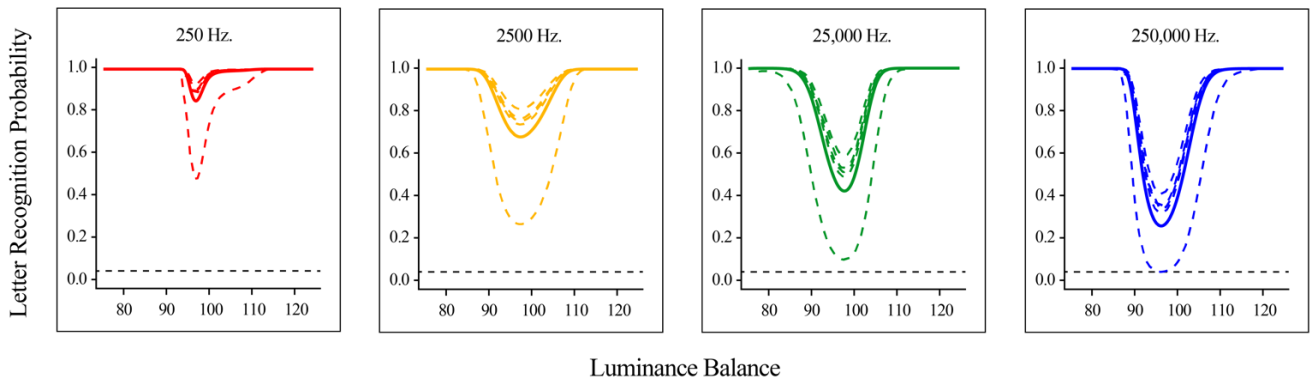

Figure 5. Regression models for respondents 11-15 reflect the judgments from Treatment Set 3, wherein 2-microsecond flashes were delivered at frequencies ranging from 250 to 2000 hz The minimum probability levels were all well above chance, and became progressively higher as frequency was reduced. .

*Treatment Set 4*    The findings reported above demonstrate that a 250 hz frequency in combination with a 2-microsecond flash duration is especially effective at enhancing luminance contrast of the flicker-fused letter. To better evaluate the role of duration, five new respondents were tested with displays using flash durations at 2, 20, 200, and 2000 microseconds, each at 250 hz. The logistic regression models are provided in Fig 6. One can see that the recognition minimum moved progressively higher as flash duration was decreased.

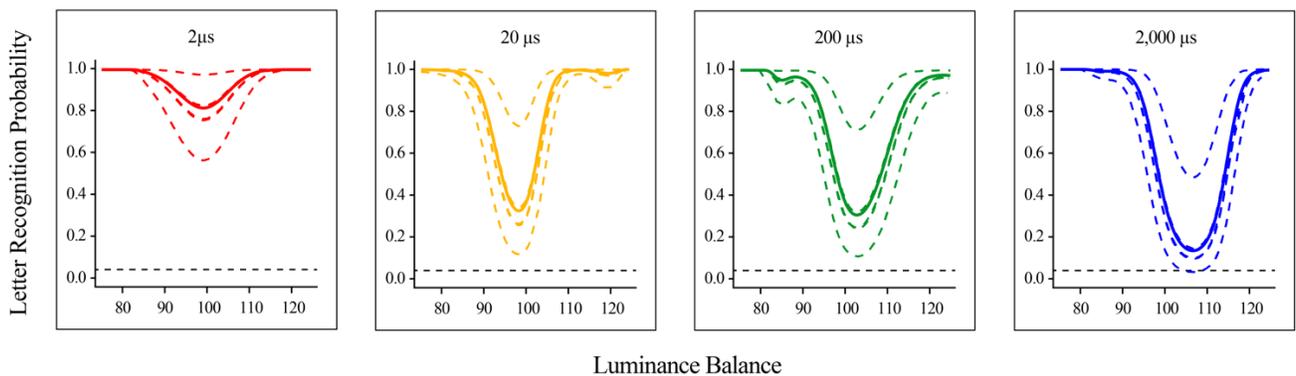

Figure 6. Respondents 16-20 judged the displays of Treatment Set 4, which varied the duration of flashes from 2 to 2000 microseconds, each level being shown at 250 hz. Minimum probability of recognition was only moderately above chance for the 2000-microsecond combination. The minimum became progressively higher as flash duration was reduced.

*Treatment Set 5*    A 2-microsecond flash provides luminance that matches the background when the frequency is 250,000 and the duty cycle is 50%. For this combination the intensity of the flash is only twice the level of background intensity.  However, as frequency is reduced, the intensity of a 2-microsecond flash must increase to maintain a luminance that matches the background.  In other words, reductions in frequency from 250,000 hz requires a progressively longer period, with the 2-microsecond flash duration providing light for a progressively smaller portion of that period.  The aggregate amount of light being delivered by the flicker-fused stimulus must remain the same, so the flash must become more intense.

Treatment Set 5 addressed whether the violation of the Talbot-Plateau law was due to flash intensity.  For this task we specified the bright and dark portions of the flash cycle as departures from background intensity, and for convenience, describe them as bright flashes and dark flashes.  Bright/dark flash pairs were displayed for a variable number of cycles ranging from 1 to 128, the duration of each cycle matching the period of a 250 hz frequency, i.e., 4 milliseconds.  Bright/dark flash amplitudes were at 100% and 50%, wherein 100% matched the conditions provided by a 250 hz x 2 microsecond combination. The 50% bright and dark flashes were at half those amplitudes.

Logistic regression models for these treatment conditions are provided in Fig 7.  They show that recognition was near zero for letters displayed for one or two cycles (8 ms).  The probability then increased as a linear function, reaching reliable (100%) letter recognition for most of the respondents by eight cycles (32 ms).  The regression models for 50% and 100% flash intensities were essentially identical.

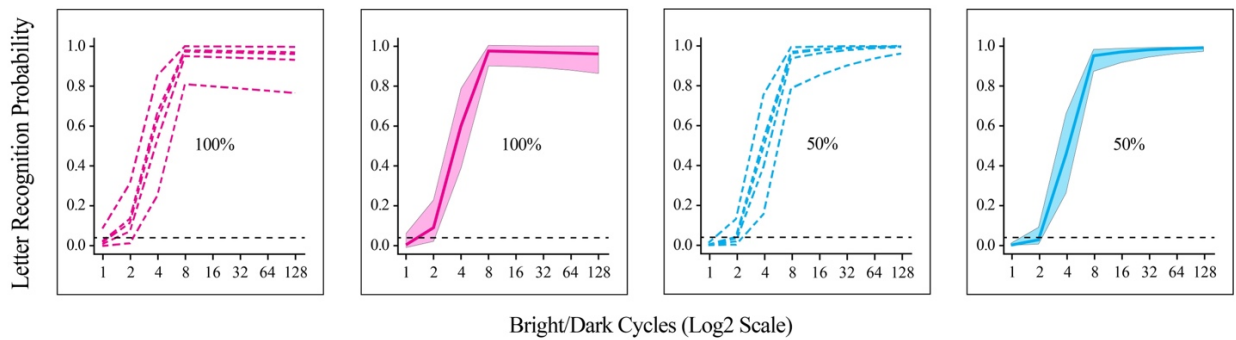

Figure 7.  Treatment Set 5 tested respondents 21-25 with luminance-balanced flash sequences, with bright and dark amplitudes being specified as departures from background level.  For the 100% condition the bright and dark components of each flash cycle had the same amplitudes as  (250 hz x 2 microsecond) displays.  Bright and dark components of the 50% condition were at half amplitude. Each trial displayed a number of cycles ranging from 1 to 128.

It should be emphasized that for 100% and 50% amplitudes, luminance of the flicker-fused letter is equal to background irrespective of the number of cycles being displayed.  The number of cycles is providing evidence that there is a cumulative build-up of activity that provides for the contrast enhancement.  The fact that the models at 50% and 100% intensity are virtually identical provides evidence that the perceived contrast is not due to a physical imbalance of luminance.  Instead, we infer that the extreme ratio of bright and dark durations serves as a contrast trigger, registering the flicker-fused letter as having a luminance level that differs from background.  The bright and dark portions of the cycle are no longer able to combine their influence on retinal mechanisms, and instead one or the other takes control of perception.

*Comparing Treatment Sets*     Overall, there was substantial consistency of effect from the various treatment combinations.  Each treatment combination, except (250 hz x 2 microseconds of Fig 3), provided a minimum probability of recognition that was within 10% of the Talbot-Plateau prediction (100%T-P).  Range for producing brightness-match judgments was no greater than 10% above and below the minimum for most respondents across all treatment combinations, again with the exception of (250 hz x 2-microsecond) of Fig 3.  Minimum recognition became progressively higher as frequency was reduced from 25,000 hz to 2500 hz (Fig 4), with further increases in recognition probability as frequency dropped across four frequency levels from 2000 hz to 250 hz (Fig 5).   In a similar manner, the minimum probability of recognition increased as flash durations, displayed at 250 hz, decreased from 2000 to 2 microseconds (Fig 6, with Figs 4 and 5 providing anchoring data in the 250 hz panel).

Each of the frequencies in Treatment Set 3 provided peak minimum recognition that was below 100%T-P.  Almost all the other treatment combinations provided peaks that were above 100%T-P, though Treatment Set 1 did find it to be below the expected value at 2500 hz.  Respondents were extremely consistent in the location of the minimum peak for each of the treatment sets. so the peak shift does not reflect an anomalous test session.  Future work will be needed to pin down the source(s) of these performance shifts.

We appear to be finding frequency and flash duration ranges wherein flicker-fused letters that are luminance-matched to background can be registered by the visual system.  The probability of recognition becomes more reliable as the frequency of the stimulus approaches 250 hz and the duration is reduced to 2 microseconds.  The letter is perceived as being at a luminance that differs from the background, so we infer the activation of mechanisms that would register stimulus contrast.  The finding that half-amplitude bright/dark departures from background provide letters that are recognized as well as full-amplitude departures suggests that an imbalance of the duration of bright and dark components of the flash sequence is what triggers detection of contrast.

## DISCUSSION
### *Where Talbot-Plateau Correctly Predicts Brightness Match (Or Doesn't)*

The Talbot-Plateau law was found to be valid for the conditions provided by Treatment Set 1, where frequency ranged from 250 hz to 250,000 hz and where a 50% duty cycle was used for each of the display frequencies.  The task varied luminance of the flicker-fused letters relative to background luminance.  When flash duration, frequency, and intensity of the letter provided luminance at about 10% below the background, a dark letter was seen. When flicker combinations provided luminance that was about 10% above the background, a bright letter was seen.  The expectation was that the letter would not be seen for Treatment Sets where the Talbot-Plateau law predicted equal luminance for letters and background.  The letter would simply be invisible and the respondent would be unable to identify which letter had been displayed. Results of testing with the first protocol supported these predictions (see Fig 3).

The physiological mechanism for registering the luminance of flicker fused stimuli needs to be discussed.  It begins with photoreceptors passing their responses to ON and OFF channels (Gotch (1903; Einthoven & Jolly, 1909; Hartline, 1928).  A number of investigators have suggested that when the frequency of the flash sequence is above the fusion threshold, transient-responding channels change to provide sustained response (Enroth, 1952; 1953; Granit, 1955; Arduini and Pinneo, 1962; Arduini, 1963; Pinneo and Heath, 1967).  Some have said that the OFF channels "lose their nature" once they transition to a steady firing rate.  However, as noted by Burkhardt & Fahey (1999), in addition to providing a transient response, the ON- and OFF-responding channels provide a bi-directional change in their steady firing rate.  ON cells increase their firing rate with an increase in light level and decrease it when the light level drops.  OFF cells respond with the opposite response to a change in light level.  For a steady light at the upper end of the response range, the ON cells are firing at a high rate and the OFF cells are firing at a low rate.  The reverse is true for steady light at the low end of the response range.  Luminance level of a stimulus is being registered by a dual-coding system that one might presume serves to more precisely deliver log-linear Weber levels by counterbalancing anomalies that one would expect from biological systems.

The ON and OFF components of the dual-channel system carry the luminance information to other parts of the visual system. DeValois and associates (1962) found the same increment/decrement balance with extracellular recordings fr macaque lateral geniculate nuclei. There was an initial increase or decrease in firing rate when an adapting light was initially turned on, which undoubtedly reflected relay of transient ON and OFF retinal activity. But then the relay cells returned to a steady level of discharge, with ON channels firing at a higher rate than before and the reverse for OFF channels.

Bartlett & Doty (1974) found that about half the neurons in primary visual cortex in squirrel monkey responded in this way, and similar results were reported by Kayama and associates (1979). The neurons provided transient ON and OFF responses to the initial onset of a diffuse light, but then continued to respond at a steady level as a function of luminance level across at least a three-log range. Peng & Van Essen (2005) provided further support for the dual-channel hypothesis based on neuron activity in V1 as well as V2. Various neuron responded best to a limited range of luminance, which these authors suggested might provide for discrimination of specific gray levels.

We previously found that the Talbot-Plateau law provided robust and relatively valid predictions about what treatments would yield equal brightness judgments for flicker-fused and steady stimuli (Greene & Morrison, 2023). That work was done with the flicker-fused and steady stimuli being judged against a dark background. The results from Treatment Set 1 extends support for the Talbot-Plateau law, finding that it holds for a luminance-match discrimination task across a large range of frequencies as long as the duty cycle is at 50%, i.e., equal ON and OFF durations.

Although the Talbot-Plateau law is presented as a theory of brightness perception, it should be clear that it is specifying the physical luminance levels of the stimuli. Flash sequences that provide a match in brightness of flicker-fused and steady stimuli are producing equal physical luminance for the two. For convenience we have described these stimuli as being "luminance balanced."

The Talbot-Plateau predictions were not found to be valid where the flicker-fused letters were displayed at flash frequencies in the vicinity of 250 hz in combination with ultra-brief flash durations. The flicker-fused letters and the background were luminance balanced, yet contrast was perceived. What would cause the visual system to register the letter as having a different luminance level?

*Balance of ON/OFF Durations Can Affect Retinal Channel Selection*

The question at hand is how a luminance-balanced letter could trigger the perception of luminance contrast. What seems most plausible is that stimulus frequency is itself interacting with the intrinsic signaling mechanisms of the ON and OFF channels to generate the contrast signal. This would be similar to how the flicker of black and white sectors on a spinning disk can induce Fechner colors (Fechner, 1838; Benham, 1895; von Helmholtz, 1924; Cohen & Gordon, 1945). Unlike natural colors, the Fechner colors (also known as Benham colors, and flicker colors) are not based on differential wavelength absorption of the cones. Rather, they are thought to be due to frequency-based activation of color-encoding retinal channels (Pieron, 1922; Fry, 1933; Campenhausen, 1968; Festinger et al., 1971; Grunfeld & Spitzer, 1995). The corresponding concept that we are adopting here is that the frequencies that yield perception of the flicker-fused letters are activating endogenous ON and OFF signaling mechanisms that provide the letters with a perceptible difference in luminance, i.e., contrast.

For the flicker colors, the hue that is elicited is a function of frequency. One of the earliest stimulus configurations used by Fechner (1838) was a simple half-white, half-black disk. Spinning the disk can produce gray, but also yellow or blue depending on the speed of rotation. Most of the quantitative work on flicker colors has used variations of the Benham disk, which is illustrated in Figure 8. When the disk is spun clockwise at about 5-10 hz, the thin-line arcs labeled as A will circle around and are seen as a blue ring, the ring from the B arc appears green, and the C arc produces a red ring. If one spins the disk in the opposite direction, the colors in the A and C rings are reversed. With this configuration the colors are relatively unsaturated. Color disappears if the lines of the arc are replaced by black strip, but flicker colors can be elicited with a gray strip (Festinger et al., 1971).

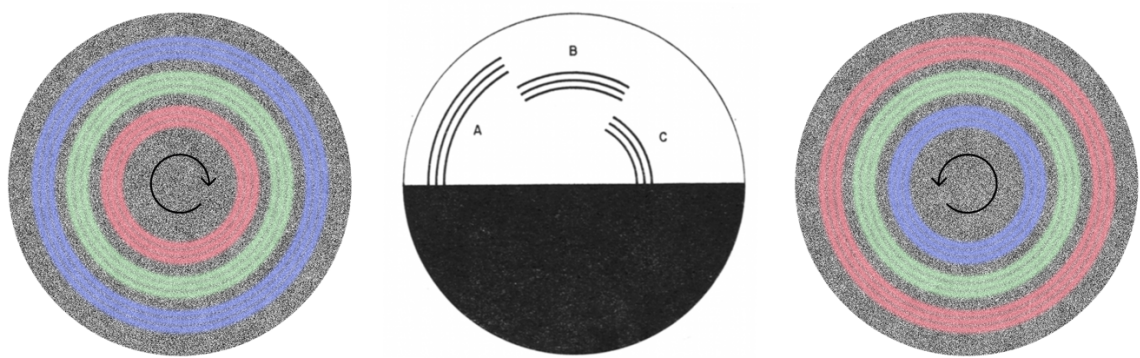

Figure 8.  The Benham Disk adds arc lines to the basic black/white disk.  The black and white halves generate a flickering gray background, and the line arcs elicit the perception of color.   It is thought that the black and white flicker is able to selectively activate retinal color channels, depending on the timing of black and white stimulation.

Both & Campenhausen (1978) proposed that the activation of color channels is determined by the phase of ON and OFF activation.  This view received support from Tritsch (1992) who replaced the black and white zones and a single arc strip with differential texture density.  This eliminated high Fourier components from the stimulus, yet the spinning disk could still elicit perception of reddish-brown and blue..  He conceded that producing flicker-green would require adding higher Fourier components.   Schramme (1992) made the hypothesis more explicit by claiming that the fundamental sinusoidal stimulation by the black and white zones selectively activates blue/yellow receptive fields of retinal ganglion cells.  This would be consistent with the view that the basic color vision circuity of mammals is dichromatic, with the trichromatic mechanisms of old-world primates being added as a recent evolutionary development.

The spatial structure of color-inducing arcs on the Benham disk appears to be relevant.  Campenhausen (1968) reported that the arc-paths of the Benham disk lose their color if the lines are made too thick.  This is interpreted as modifying the balance of center/surround receptive fields.  Grunfeld & Spitzer (1995) endorsed this view by incorporating the center/surround response profiles of color-responding ganglion cells in their flicker-color models.

Although one can generate perception of blue and yellow colors as a function of ON/OFF frequency, it appears that a more complex temporal pattern is needed to produce most hues.  Festinger and associates (1971) adjusted positioning and length of the Benham arcs, creating ON and OFF pulse shapes that elicited at least six different hues (red, blue, yellow, cyan, magenta, green) plus neutral gray. Courtney & Buchsbaum (1991) derived models based on the impulse responses of wavelength-activated channels, and found them to be a fairly close match to the profiles proposed by Festinger and associates (1971).

The frequencies that produce flicker colors are very low, e.g., 5-10 hz, well below frequencies that were used here to produce enhanced luminance contrast. However, the low fundamental frequency may not preclude precise timing within the color channels. Both and Campenhausen (1977) reported that a small difference in length for two adjacent arcs on a Benham disk can affect the flicker colors. A perceptible difference in color and brightness was found with stimulus durations that differed by only 50 microseconds.

At this point it is unclear whether a more complete model of color processing by the retina would explain all the flicker-color phenomena. But whatever might be the details, the phenomenon itself makes it plausible that a luminance-balanced stimulus could generate contrast through anomalous activation of retinal neurons. Observers see the luminance-balanced letter when the ON activation is very brief and the OFF activation is relatively long, e.g., 2 microseconds and 4 milliseconds, respectively. The letter is seen whether the ON/OFF amplitudes are at their limit or are at half amplitude (see Fig 7), so the imbalance of pulse duration appears to be what triggers luminance contrast.

### *Retinal and Optic Nerve Oscillations Convey Luminance and Contrast Signals*

Oscillations may play a role in signaling luminance and contrast, and the successive ON and OFF activation of flicker stimuli may well recruit oscillations among retinal neurons. Retinal oscillations have been documented across a large range of vertebrate and invertebrate species. Frohlich (1914) was the first to note oscillatory potentials in the visual system, with onset of illumination producing 20-90 hz sinusoids in the electroretinogram of octopus, and offset providing 20-40 hz. Adrian & Matthews (1928) found similar responses in the optic nerve of eels, with frequency of the oscillations being a function of illumination level. Granit (1933) was the first to observe the oscillations in mammals. The ON and OFF light transitions produced 100-150 hz oscillations recorded in the optic nerves of cats, which he attributed to synchronized action potentials. Change in light intensity affected the amplitude of the response but not the frequency. Noell (1951; 1953) found similar responses in pigeon, rabbit, and monkey.

Most investigators report low-frequency oscillations with the onset or offset of stimuli, but there have been numerous examples at relatively high frequencies. Electroretinogram activity in the 140-160 hz range has been reported in humans (Cobb & Morton, 1954; Bornschein & Goodman (1957). Doty & Kimura (1963) found that a flash of light evokes oscillations in the optic nerve of monkeys ranging up to 160 hz. Again, frequency was independent of intensity. They suggested that ganglion-cell oscillations were driving the production of action potentials that then propagated to V1 where oscillations above 200 hz were found.

Laufer & Verveanu (1967) observed that the onset of a steady stimulus produced regular oscillations in the optic tract of cats that followed an initial "impulse response." These were observed with a gross electrode that were registering spike clusters at a frequency of 40-100 hz, again irrespective of illumination from 0.015 to 90 lux. Simultaneous recording from the retina and optic nerve suggested that the clusters were coming from synchronized firing of several retinal neurons, presumably the ganglion cells. They cited several studies that claimed no intrinsic oscillations from individual cells, so these investigators suggested coordinated circuit activity.

As noted above, several laboratories have found that the frequency of spikes recorded from the optic nerve is not much changed as a function of the level of illumination (Adrian & Matthews, 1928; Granit, 1933; Doty & Kimura, 1963; Laufer & Verveanu (1967). This has generally been attributed to adaptation, in which horizontal cells modulate the light-activated signal at the photoreceptor-bipolar synapse. However, the differentials of light level that provide for brightness judgments across a range of luminance must be delivered to cortex. If the luminance level is being specified using a rate code, one would expect some differential in firing rate, notwithstanding the role of adaptation. We submit that the lack of change in sustained frequency of spikes provides additional support for the proposition the luminance level is being transmitted using a dual code, i.e., reciprocal firing rates from ON and OFF channels. An increase in luminance level increases the firing rate of the ON channel and simultaneously decreases the firing rate of the OFF channel, with the converse happening for a decrease in luminance level.

*Intracellular Recording of Oscillations in Amacrine Cells*

Amacrine cells may be especially involved in selecting the output channels for luminance and contrast. Although there are a great many types of amacrine cells, their electrophysiological response is often characterized as ON, OFF, and ON/OFF, and also as providing transient versus sustained activity levels (Kaneko & Hashimoto, 1969; Toyada et al., 1973; Chan & Naka, 1976; Djamgoz et al., 1985; Teranishi et al., 1987). DJamgoz and assocates (1990) report that sustained responses that might relate to perception of luminance appears to be provided by OFF bipolar terminals that stratify only in the "a" intralaminar layer, with ON terminals stratifying in the "b" layer. Transient ON/OFF responses are provided by endings that stratify in both layers. The same functional stratification was suggested by Sakai & Naka (1990a) and Ammermuller & Kolb (1995).

Sakai & Naka (1990a) did simultaneous intracellular recording in amacrine and ganglion cells of catfish. A 10 millisecond flash produced a sharp depolarization that could be recorded in both cell types, which was followed by a dampened oscillation at about 35 hz. Sakai & Naka (1990b) reported that spontaneous oscillations of the amacrine cells were at the same frequency as the damped oscillations produced by a flash. Mutual interaction between amacrine and ganglion cells could be elicited by injection of a current pulse or white-noise, with stimulation of one of the cells producing damped oscillations in the other (Sakai & Naka, 1990a). However, this only occurred if the cells were of the same polarity.

Burkhardt & Fahey (1999) provided intracellular recordings from salamander ON/OFF amacrine cells, registering response amplitudes as departures from a background luminance level. The ON-OFF amacrines manifested great contrast sensitivity, with up to a third of the cells responding to a contrast differential of only 1%. A contrast step of about 1% is just detectable when displayed to the human fovea under favorable conditions (Burkhardt et al., 1987; Westheimer et al., 1999). The amount of contrast required to elicit a half-maximum response for one contrast polarity was generally less than 10% for almost all the cells, and half the cells responded to either polarity at that level of sensitivity. The amacrine population varied greatly in the size and balance of ON vs OFF responsiveness, which suggested a method for distributed encoding of luminance contrast (Burkhardt & Fahey, 1999).

Rod bipolar cells of rat control transient and sustained activity in AII amacrines, with the amplitude of the sustained response reflecting absolute luminance level, and with the transient response to light steps having the precision of Weber's law (Oesch & Diamond, 2011). Similar results were reported by Graydon and associates (2018). Oesch & Diamond (2019) suggested that the responses of A17 amacrines "extended the range" of contrast computation.

Solessio et al. (2002) reported intrinsic oscillatory responses in wide-field amacrine cells of white bass, which modulated the ON and OFF activity of ganglion cells. These cells were bistratified, which likely reflects transmission of ON/OFF responses. Stimulation with a depolarizing current induced oscillations. If administered when the cell was depolarized, it produced a transient oscillation that lasted for about 40 milliseconds. If stimulated when hyperpolarized, it generated a stable oscillation that lasted for the duration of the stimulation. Oscillation frequency increased logarithmically as a function of mean membrane potential, ranging as high as 140 hz for the steady oscillations and 300 hz for the transient oscillations. The oscillation generator was found to be intrinsic to the cell, being created by the interplay between voltage-gated calcium and potassium currents. Follow-up work on amacrine cells that had been isolated from the retina concluded that a membrane resonance of ~100 hz was based on interplay between voltage gated L-type $Ca^{2+}$ channels and calcium-dependent $K^+$ channels (Vigh et al, 2003).

It bears repeating that the amacrine-cell oscillations are thought to modulate the activity of ganglion cells (op cit.) One might further suggest that the frequency and balance of ON/OFF activity could select among alternative oscillators, and determine which ganglion cell is activated. The channel-selective role of the amacrine influence could register luminance and contrast, but might also have a role in transmitting shape-relevant information.

### *Oscillations and Synchrony for Encoding Stimulus Attributes*

Retinal mechanisms may have encoded specific features that made it possible to identify the flicker-fused letters. A number of investigators have suggested that registration of image content is provided by synchronized activity among a distributed population of neurons (Miller, 1974; von der Malsburg, 1985; Singer & Gray, 1995). Similar functional linkage of ganglion cells in mouse were reported by Roy et al. (2017). Oscillatory potentials appear to be critical in providing for the synchronized activity, and robust oscillations have been noted in both the retina and lateral geniculate nucleus (Doty & Kimura, 1963; Fuster et al., 1965; Laufer & Verzeano, 1967; Arnett, 1975; Ariel et al., 1983; Munemori et al., 1984; Ghose & Freeman, 1992).

Neuenschwander & Singer (1996) reported that oscillatory activity in retinal ganglion cells provides for synchronized activity for stimuli that were separated by up to 20 degrees of visual angle.  This was characterized as "binding" the separate stimulus elements into a unitary percept.  The oscillatory activity ranged from 61 to 114 hz.  The oscillations were not phase-locked to stimulus onset and could be triggered by stationary as well as moving stimuli.  This temporal structure was reliably transmitted to the lateral geniculate nucleus, which suggests that it was serving as a behaviorally relevant signal.  Castelo-Branco et al. (1998) simultaneously recorded from retina, lateral geniculate nucleus, and visual area 18, finding reliable transmission of oscillations and synchronized firing, with static stimuli being more effective than moving stimuli.

Neuenschwander and associates have provided a comprehensive discussion of the role of oscillations in synchronizing retinal, thalamic, and cortical activity, and its potential for encoding stimulus features (Munk & Neuenschwander, 2000;  Neuenschwander et al. (2002).  The ability to identify briefly displayed letters could depend on synchronized oscillations of spatially separated receptive fields.  Those oscillations might be recruited through resonance that is driven by a flicker-fused stimulus.  But as was found with flicker colors, eliciting a specific stimulus feature could depend not only on the frequency of ON/OFF stimulation, but also temporal patterning within each cycle.

**CODA**

Contrast discrimination has proved to be an effective tool for evaluating the Talbot-Plateau law.  Displaying a flicker-fused letter embedded within a surrounding background allows one to test its predictions across a larger range of alternative treatments.  Requiring letter recognition reduces the probability of chance performance to 4%, which provides a greater range for quantifying strength of treatment influence.  For the present treatments that used a 50% duty cycle for specifying flash frequency and duration, the Talbot-Plateau principle correctly predicted the intensity that was needed to match the luminance level of flicker-fused letters to the steady background.  These results support the pioneering work that was done in the early part of the 20th century that affirmed the Talbot-Plateau proposal as a law.

We have found, however, that there are limits to the law.  It failed to predict brightness match when the frequency of the flicker-fused stimulus was in the range between 250 and 2500 hz and the duty cycle of the flash sequence was exceptionally low. Flicker-fused letters were identified when their physical luminance, based on the average intensity of the flash sequence, was equal to background luminance. We believe that an imbalance in activation of ON and OFF cells in the retina produced anomalous selection of contrast-encoding channels, or possibly channels that conveyed shape-specific attributes. Given that the Talbot-Plateau law does not apply under these conditions, the findings suggest exciting new visual mechanisms that need to be explored.

**Acknowledgments.** Alexander Kezar conducted the experimental test sessions. Statistical modeling was done by Dr. Wei Wang, Harvard Medical School. Figure graphics were rendered by Rob Strong. This research was supported by the Neuropsychology Foundation and the Quest for Truth Foundation.